# High-Harmonic Optical Vortex Generation from a Plasma Aperture

Runze Li,[1] Wenchao Yan,[1, 2,*] and Longqing Yi[1, 2,*]

[1]*State Key Laboratory of Dark Matter Physics, Key Laboratory for Laser Plasmas (MoE), School of Physics and Astronomy & Tsung-Dao Lee Institute, Shanghai Jiao Tong University, Shanghai 200240, China*

[2]*Collaborative Innovation Center of IFSA (CICIFSA), Shanghai Jiao Tong University, Shanghai 200240, China*

*e-mail: wenchaoyan@sjtu.edu.cn; lqyi@sjtu.edu.cn

## ABSTRACT

When a high-power, femtosecond, circularly polarized (CP) laser pulse is incident on a micrometer-scale aperture in a solid foil target, it drives surface plasma oscillation, generating high-order harmonic vortices in the diffracted light[1]. However, this mechanism has so far only been studied theoretically under ideal conditions. In this work, we perform numerical studies on more realistic situations. In particular, we focus on a scenario where the laser is obliquely incident on the target surface to avoid the potential damage of the optics by the reflected light. We demonstrate that increasing oblique incidence angle, reducing target thickness, and improving laser contrast can enhance the harmonic conversion efficiency. However, the generated harmonic beams may contain both Laguerre-Gaussian (LG) (vortex) and non-LG components under non-ideal conditions. We show that they can be separated by their divergence, as the vortex components has smaller diverging angle. In addition, we have performed computational analyses on the harmonic divergence angles and topological charge spectra of vortex high-order harmonics under different conditions. These high-order harmonic pure LG modes can potentially be filtered out for wide range of fundamental and applied physics researches. This study provides valuable insights for the design and implementation of future experiments.

## I. INTRODUCTION

Light carries both spin and orbital angular momentum (SAM and OAM)[2–4]. The SAM is associated with the polarization direction of CP light ($\pm\hbar$ per photon)[5]. The OAM is possessed by light beams with helical wavefronts. The helical profile is expressed by a helical spatial phase $exp(il\phi)$, where $l$ denotes the topological charge, and the $\phi$ represents the azimuthal angle, indicating that each photon carries an OAM equal to $l\hbar$[6]. Recently, vortex high-order harmonics have attracted considerable attention due to their promising applications in diverse fields, including optical communication[7–10], optical trapping[11], and biophotonics[12].

Currently, extensive research is focused on generating high-power, high-frequency vortex laser beams[13–18], which are of significant interest for probing and topologically controlling ultrafast light-matter interactions[19,20]. Under relativistically intense laser conditions, most of the proposed methods for producing such beams rely on high-order harmonic generation (HHG) resulting from the interaction between high-power lasers and solid targets. The driver can be a relativistic vortex laser pulse[21,22], a CP pulse[23,24] or a linearly polarized laser beam[20]. However, these approaches are all based on the relativistic oscillating mirror (ROM) mechanism[25–27], which is suppressed for CP drivers under normal incidence[28]. Generating high-intensity, high-frequency CP vortex beams remains challenging, yet such beams are of particular interest for controlling chiral structures and optical manipulation at relativistic intensities[29–31].

Recently, we proposed a mechanism termed the relativistic oscillating window[1], in which vortex high-order harmonics are generated behind the solid target when a CP laser diffracts through a circular aperture in the target. In this process, the laser drives chiral electron oscillation at the aperture periphery, generating high-harmonic optical vortices via the Doppler effect and spin-orbit interaction of light. Compared with the ROM mechanism, the key advantage of our diffraction-based scheme lies in its intrinsically two-dimensional (2D) electron dynamics within the diffraction screen. In contrast, electron motion in the ROM process is predominantly one-dimensional (1D), occurring perpendicular to the target surface[32–34]. This fundamental difference provides an additional degree of freedom for controlling laser-plasma interactions and tailoring the optical properties of the emitted harmonic beams.

However, this mechanism has thus far been studied only theoretically under some ideal scenarios, such as normal incidence, perfect alignment, negligibly thin target, and limited pre-plasma expansion. In response to the specific requirements of the experiment, we have conducted an in-depth numerical investigation into this process. To avoid potential laser damage from back reflected light—a common risk in normal-incidence experiments—we consider a scenario where the laser is obliquely incident on the target surface while propagating parallel to the axis of a pre-drilled plasma aperture. In addition, we investigate the effects of laser and plasma parameters, such as target thickness, pre-plasma scale length due to limited laser contrast, as well as the influence of the laser incident angle.

This work is organized as follows: section II presents our numerical simulations of the process, validating the spin-orbit angular momentum interaction (SOI) by analyzing the harmonic phase distribution in the transverse cross-section. Furthermore, we compute the angular distributions of harmonics at different orders. In Section III, the high-harmonic energy spectra are calculated under various conditions. These results provide valuable insights for enhancing the harmonic conversion efficiency in experiments. Additionally, given the practical requirement of harmonic collection in experiments, we conducted calculations and analyses on the angular distribution of high-order harmonics. In Section IV, the topological charge spectrum of high-order harmonics is computed and analyzed using the azimuthal Fourier transform and its inverse. We further separate the LG harmonic component from it and compute its angular distribution, while it is noteworthy that such isolation of the LG mode has been demonstrated to be experimentally feasible. These analyses are essential for experimentally characterizing the vortex properties of the generated harmonics.

## II. Simulation setup and main results

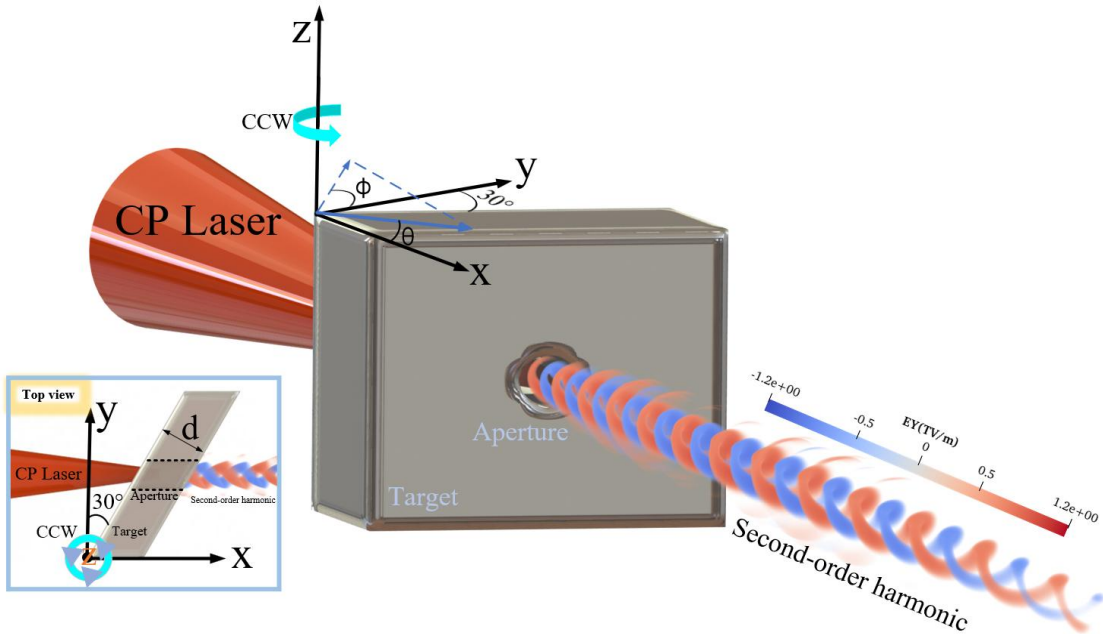

FIG.1 A high-power CP laser is incident on the exact center of the aperture in a thin-film target, driving electron oscillation at the periphery and generating vortex high-order harmonics in the diffracted light. The laser propagates along the X-axis, the inner wall of the target hole is also aligned along the X-axis, and the target surface forms a 30° angle with the Y-axis (as shown in the inset at the lower left corner).

We first demonstrate our scheme using three-dimensional (3D) particle-in-cell (PIC) simulations with the code EPOCH[35]. The setup of laser-target interaction is illustrated in Fig. 1. A CP laser beam with wavelength $\lambda_L = 800nm$ propagates along the x-axis and is focused onto the center of the target's front surface. The laser-pulse spatial is defined in terms of intensity as Gaussian and temporal is $sin^2(\pi t / \tau_0)$, where $0 < t < \tau_0 = 60fs$ [pulse duration $\tau_L = 30fs$ (full width at half maximum)]. The intensity of the laser beam is $I_0 \approx 3.4 \times 10^{19} W / cm^2$, corresponding to a normalized amplitude $a_0 = eE_0 / mc\omega_0 \approx 2.8$, where $e$, $m_e$, $c$ and $\omega_0$ denote the elementary charge, electron mass, velocity of light in vacuum, and the laser frequency, respectively. The simulation box has dimension of $L_x \times L_y \times L_z = 46\mu m \times 30\mu m \times 30\mu m$ and is sampled by $2304 \times 640 \times 640$ cells, with four macroparticles for electrons and two for $Al^{13+}$ per cell. The target is composed of pre-ionized plasma with a thickness of $d = 10\mu m$ and an electron density of $30n_c$, where $n_c = m_e\omega_0^2 / 4\pi e^2$ is the critical density. The radius of the aperture is $r_A = 2.5\mu m$, with a density gradient at the inner boundary $n(r) = n_0 \exp[(r - r_A) / h]$ for $r < r_A$, where $h = 0.2\lambda_L$ is the scale length. A 30° angle is set between the target surface and the plane perpendicular to the laser propagation direction to prevent damage to the laser system from back-reflected light in practical experiments. Crucially, the plasma is pre-drilled such that the inner surface of the aperture remains aligned parallel to the laser beam axis (as illustrated in the top view in the inset).

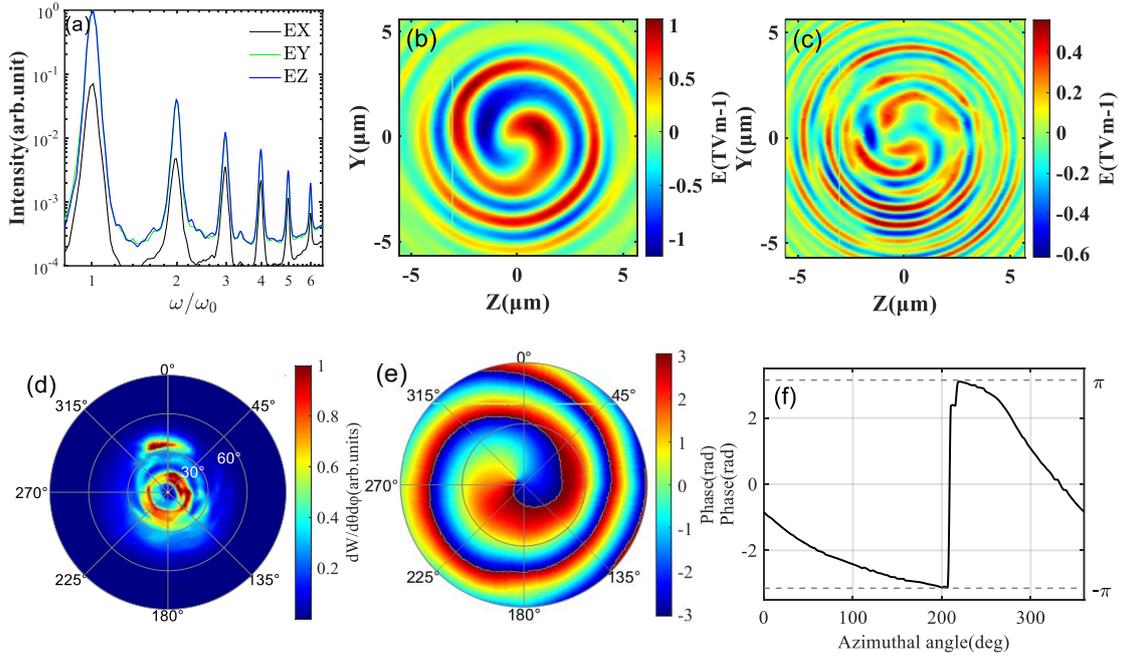

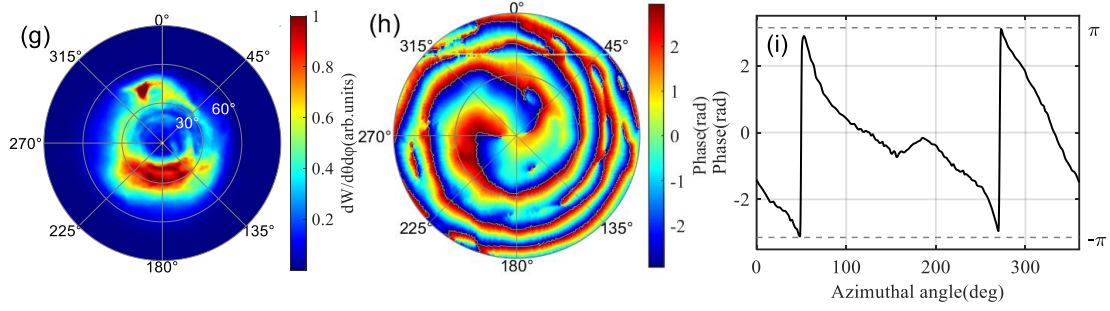

FIG.2 (a) Harmonic Spectrum of Diffracted Electric Field $E_x$, $E_y$ and $E_z$. (b) and (c) show the harmonic Ey fields distribution in the transverse cross-section with frequency $2\omega_0$ and $3\omega_0$, respectively. (d-e) shows the angular intensity distribution in the forward direction and the phase of second harmonic, and (f) is the 1D phase distribution along $r = 0.61 \mu m$ in (e). (g-i) are the same as (d-f), but for the third harmonic beam.

The energy spectra of the electric field components, obtained via Fourier transform of the diffracted light, are shown in Fig.2(a). It is evident from the energy spectrum that both odd-order and even-order harmonics coexist. By filtering the diffracted optical field within the frequency range from $1.5\omega_0$ to $2.5\omega_0$, one obtains the electric field distribution of the second harmonic beam, as shown in Fig.2(b). Similarly, the field distribution of the third harmonic, obtained using the same method, is presented in Fig. 2(c). The phase distributions of the second harmonic and third harmonic obtained through inverse Fourier transformation are shown in Fig.2(e) and (h), respectively. Taking the vortex center as the origin, the phase distributions along the circumferences with radii of $r_2 = 0.61 \mu m$ and $r_3 = 1.17 \mu m$ are shown in Fig.2(f) and (i), respectively. One can see the optical phase change $2\pi$ and $4\pi$ for the second and third harmonics within $2\pi$ range of the azimuthal angle, indicating they are vortex beams with topological charges of $l_2 = 1$ and $l_3 = 2$, respectively. This result confirms the occurrence of SOI during the interaction between the laser and the aperture target. The n-th harmonic carries an OAM of $|l_n| = n - 1$.

For future experimental studies, it is crucial to estimate the divergence of the harmonics in order to collect and diagnose HHG vortex beams. This is determined by the electric and magnetic radiation fields $(E_x, E_y, E_z, H_x, H_y$ and $H_z)$[36]. The Umov-Poynting vector:

$$\vec{S} = S_x + S_y + S_z = \vec{E} \times \vec{H}$$

denotes the magnitude of energy along the propagation direction of the harmonics, while $S_x, S_y, S_z$ represent the energy flux magnitudes in the directions of $x, y, z$, respectively. The angular distribution of high-order harmonics can be expressed in terms of $\theta$ and $\varphi$, where:

$$\theta = arccos(S_x / S)$$

$$\varphi = arctan(S_z / S_y)$$

Figure 2(d) and (g) show the angular distributions of the second harmonic and third harmonic, respectively. A pronounced increase in the divergence angle $\theta$ is observed with increasing harmonic order. Specifically, the second harmonic exhibits a divergence angle of approximately 15°, compared to about 30° for the third harmonic. It should be noted the intensity pattern presented in Fig. 2(d) and (g) deviate from the typical donut-shape of the LG pulses, this is because in the case, the harmonic beams also contain non-LG components, this will be discussed in Sec. IV.

## III. Influences of Laser-Plasma Parameters under more practical situations

The previous section presented simulations and analyses under idealized condition, where the laser irradiates at the center of aperture, propagates parallel to the middle axis, the plasma density scale length at the boundary is relatively small, etc. In real laser-plasma experiments, however, numerous factors may influence the outcomes. We therefore investigate the robustness of our HHG scheme by study the effects of a few laser-plasma parameters, including target thickness, pre-plasma, and the laser incident angle.

Figure3 presents the influence of different laser-plasma parameters on energy spectra and angular distribution. We first calculated the energy spectra for targets of varying thicknesses. As show in Fig.3(a), the harmonic intensity gradually decreases with the increase of target thickness. This is primarily because when most of the harmonic beams are generated at the front surface of the target increasing the target thickness hinders the propagation of the harmonic fields, thereby reducing harmonic intensity, this is different from the scenario discussed in [37] because here laser waist is greater than the size of the aperture. We found that to generate stronger high-order harmonics in experiments, the thickness of the target should be comparable to laser wavelength. Furthermore, the target thickness also affects the harmonic divergence angle. A comparison among Figs. 3(b-c), and 1(f) indicates that increasing the target thickness reduces the divergence angle of the harmonics within a certain range.

High-power ultrashort laser pulses are often accompanied by pre-pulses in the temporal domain, which are sufficient to ionize the target and generate pre-plasma. Figure. 3(d) presents the harmonic energy spectra for different plasma density scale lengths. One can see that as the plasma density scale length increases, the harmonic intensity drops significantly. Therefore, to enhance the high-order harmonic conversion efficiency in experiments, efforts should be made to improve the laser contrast, for instance, by introducing a plasma mirror [38–40]. The divergence angle broadens with increasing pre-plasma scale length, as demonstrated by comparing Figs. 3(e) and (f).

To investigate the influence of the laser incident angle, we simulate the scenarios in which the laser propagation direction is not parallel to the axis of aperture. Figure. 3(g) shows that when laser incidence has a certain deflection angle with respect to the axis of the plasma aperture, the intensity of harmonics gradually increases with the deflection angle. This is primarily because the interaction area between the laser and the inner walls of the target aperture increases with the oblique incidence. Additionally, the figure shows that half-integer order harmonics emerge as the deflection angle increases. This lies beyond the scope of current work and should be addressed for future studies. Figures 4(h) and (i) show the harmonic angular distribution after rotating the target counterclockwise or clockwise (top view as shown in Fig. 1) by a specific angle around the center of the aperture's incident surface. The harmonic divergence angle $\theta$ increases with the rotation angle of the target. Moreover, counterclockwise rotation enhances the harmonic intensity at $\varphi = 0°$ in the angular distribution, whereas clockwise rotation leads to increased intensity at $\varphi = 180°$.

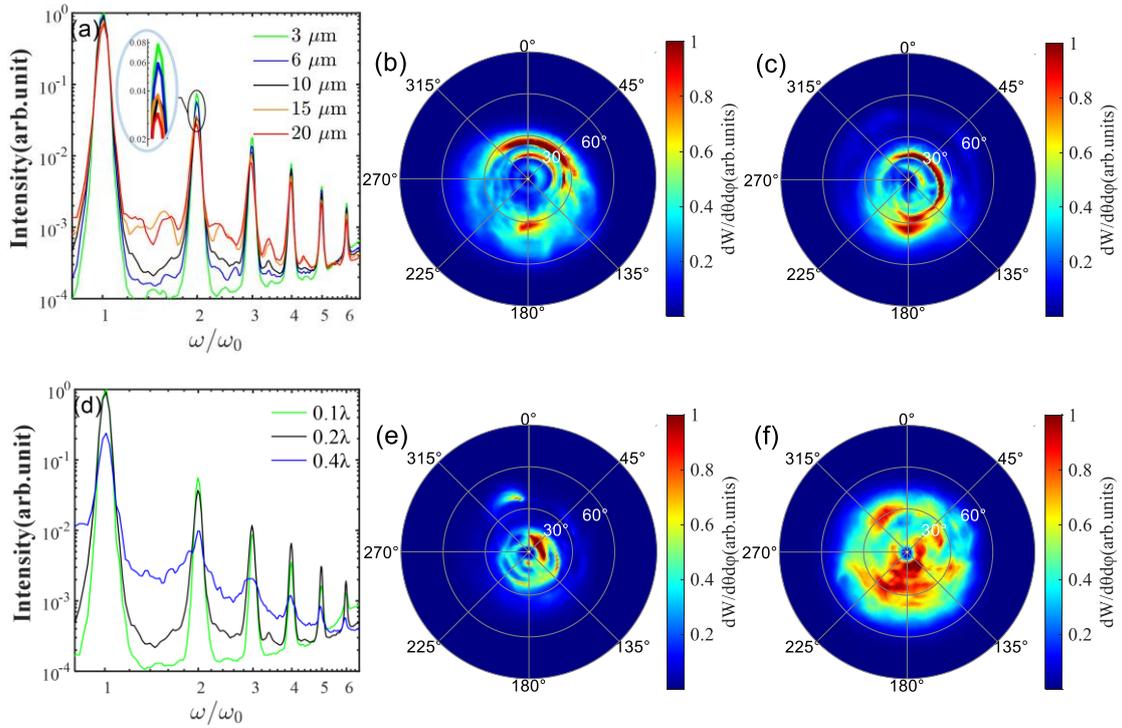

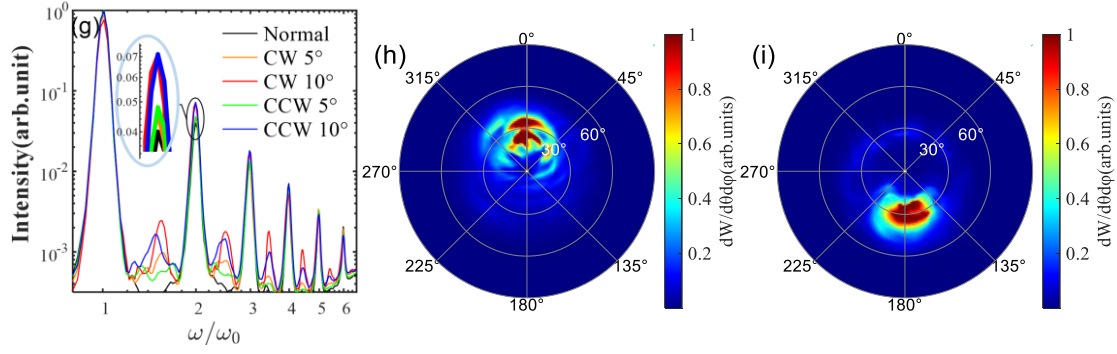

FIG.3 (a)-(c) Harmonic spectrum and angular distribution of targets with different thickness: (b) $d = 3\mu m$ (c) $d = 6\mu m$. (d)-(f) Harmonic spectrum and angular distribution of targets with different plasma density scale length: (e) $h = 0.1\lambda$ (f) $h = 0.4\lambda$. (g)-(i) Harmonic spectrum and angular distribution of targets with different deflection angle: (h) Counterclockwise 10° (i) Clockwise 10°.

## IV. Study on OAM Spectrum of HHG and Extraction of LG Components

As mentioned previously, the generated harmonics typically exhibit a complex spatial structure because they are not pure optical vortices. However, we note that the pure LG components in this mixture of electromagnetic waves can be filtered out experimentally[41,42]. To this end, it is important to compute the OAM spectrum to determine the strength of the LG components at each harmonic order. The pure vortex component is subsequently isolated from this spectrum, and its angular distribution is thoroughly investigated.

Through PIC simulations, we obtain the electric field distribution in Cartesian coordinates $E(x,y,z)$. To calculate the mode composition in high-order harmonics, we first transform the electric field distribution from Cartesian coordinates $E(x,y,z)$ to cylindrical coordinates $E(x,r,\varphi)$. The electric field can be expanded into a sum of fundamental helical modes $e^{il\varphi}$, each carrying $l\hbar$ OAM per photon[43],

$$E(x,r,\varphi) = \sum_{l=k_1}^{k_2} c_l(x,r) e^{il\varphi}$$

where $k_1$ and $k_2$ are integers representing the boundaries of the OAM spectrum of the electric field. we then perform an azimuthal Fourier transform[44] along the azimuthal angle $\varphi$.

$$c_l(x,r) = \frac{1}{2\pi} \int_0^{2\pi} d\varphi \, e^{il\varphi} E(x,r,\varphi)$$

The proportion of the electric field component with topological charge $l$ in the total electric field is:

$$P_{(l)} = \frac{1}{S}\int_0^\infty dr\, r|c_l(x,r)|^2$$

Where $S = \sum_l \int_0^\infty dr\, r|c_l(x,r)|^2$.

Figure 4(a) shows the OAM spectrum of the second harmonic. The component corresponding to topological charge $l = 1$ accounts for 58.7% of the harmonic field, indicating that the emitted harmonics are not pure vortex beams. Figure 4(b) presents the OAM spectrum of third harmonic, where the normalized weight of the azimuthal mode $l = 2$ is 21.6%. As the harmonic order $n$ increases, the normalized power $P_{(l=n-1)}$ gradually decreases. Furthermore, as shown in Figs. 4(c) and 4(d), both excessively small target thickness and overly large pre-plasma scale length reduce the vortex purity of the high-order harmonics. Therefore, to achieve high vortex conversion efficiency in experiments, the laser contrast should be enhanced and a target with an appropriate thickness should be selected.

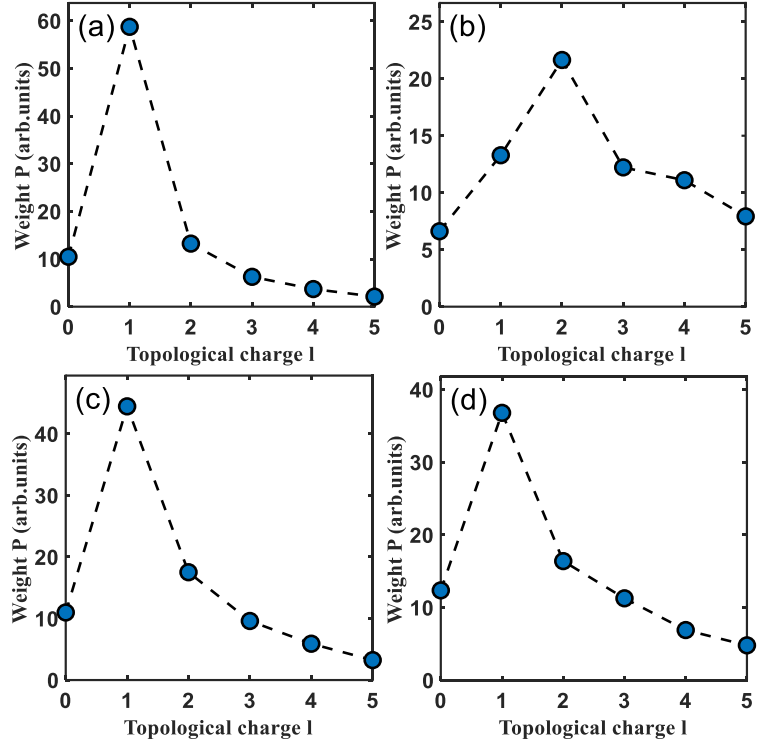

FIG.4 OAM spectrum of high-order harmonics. (a) the second harmonic (b) the third harmonic (c) The second harmonic generated by a thinner target ($d = 3\mu m$). (d) The second harmonic generated by a target with larger scale length ($h = 0.4\lambda$).

As can be inferred from the preceding analysis, the generated high-order harmonics are not pure LG beams. To extract the LG harmonic component, we perform the azimuthal Fourier transform and its inverse transform on the electromagnetic field of high-order harmonics:

$$E(x,r,\varphi) = \int_0^{2\pi} \frac{dl}{2\pi} e^{-il\varphi} c_l(x,r)$$

Using this approach, the field distribution of the LG component with topological charge $l = 1$ from the $E_y$ field of the second harmonic is reconstructed, as shown in Figs. 5(a) and 5(b). The same methodology is applied to reconstruct the electromagnetic field distributions for the topological charge $l = 1$ within the $E_z$, $B_y$, and $B_z$ components of the second harmonic, as well as for the topological charge $l = 2$ in the $E_x$ and $B_x$ components. The angular distribution of the LG mode in the second harmonic is also computed, with the results presented in Figs. 5(c) and 5(d). The divergence angle of the LG harmonic generated using a 10μm-thick target is approximately 15°, compared to about 10° for that produced with a 20μm-thick target. This demonstrates that the LG harmonic component exhibits a significantly smaller divergence angle and a more collimated beam profile. For experimental diagnosis of vortex characteristics, purer LG harmonics can be acquired by restricting the collection aperture. Moreover, increasing the target thickness appropriately facilitates the generation of LG harmonics with narrower divergence, which is advantageous for practical applications.

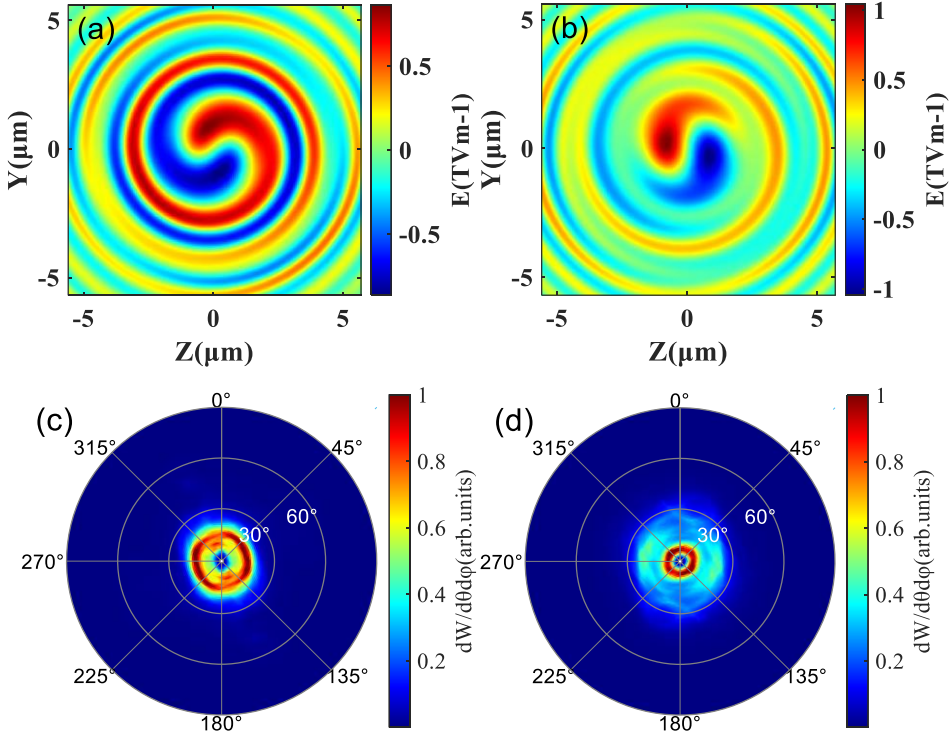

FIG.5 Top row: LG components of second harmonic field for (a)$d = 10\mu m$ (b)$d = 20\mu m$. Bottom row: Angular Distributions of LG harmonic components for (c)$d = 10\mu m$ (d)$d = 20\mu m$.

## V. CONCLUSIONS

In summary, this work investigates the interaction of a laser pulse with an aperture in a thin foil. Through analysis of the phase distribution, we demonstrate the occurrence of a spin-orbit interaction in the process, leading to the generation of vortex high-order harmonics. To account for possible experimental conditions, we further examine the effects of target thickness, laser incidence angle relative to the aperture, and pre-plasma scale length. The energy spectra of vortex harmonics under various configurations are computed, providing insight beneficial for improving harmonic conversion efficiency. Our results indicate that the conversion efficiency can be enhanced by reducing the target thickness and increasing the laser contrast. Additionally, oblique laser incidence on the aperture is found to improve efficiency, accompanied by the emergence of half-integer-order harmonics. We also analyze the harmonic angular distributions, which offers practical value for harmonic collection and experimental diagnostics. Furthermore, by employing the azimuthal Fourier transform, we resolve the topological charge spectrum of the harmonics, facilitating strategies to increase the relative weight of the LG mode. The pure LG component is subsequently isolated via the inverse transform, and its angular profile is examined, revealing a divergence angle smaller than that of the total harmonic field. Collectively, these findings provide a foundational framework for optimizing and interpreting future experiments involving vortex HHG.

**Acknowledgements**

This work is supported by the National Key R&D Program of China (2021YFA1601700), the National Natural Science Foundation of China (12475246, 12074251). The authors would like to acknowledge the sponsorship from the Yangyang Development Fund. The computations in this paper were run on the Siyuan cluster supported by the Center for High Performance Computing at Shanghai Jiao Tong University.

**Author Declarations**

**Conflict of Interest**

The authors have no conflicts to disclose.

**Data availability**

The data that supports the findings of this study are available from the corresponding author upon reasonable request.

**References**

[1] L. Yi, High-Harmonic Generation and Spin-Orbit Interaction of Light in a Relativistic Oscillating Window, Phys. Rev. Lett. 126 (2021) 134801. https://doi.org/10.1103/PhysRevLett.126.134801.

[2] J. Ni, S. Liu, Y. Chen, G. Hu, Y. Hu, W. Chen, J. Li, J. Chu, C.-W. Qiu, D. Wu, Direct Observation of Spin–Orbit Interaction of Light via Chiroptical Responses, Nano Lett. 22 (2022) 9013–9019. https://doi.org/10.1021/acs.nanolett.2c03266.

[3] Z. Shao, J. Zhu, Y. Chen, Y. Zhang, S. Yu, Spin-orbit interaction of light induced by transverse spin angular momentum engineering, Nat Commun 9 (2018) 926. https://doi.org/10.1038/s41467-018-03237-5.

[4] K.Y. Bliokh, F.J. Rodríguez-Fortuño, F. Nori, A.V. Zayats, Spin–orbit interactions of light, Nature Photon 9 (2015) 796–808. https://doi.org/10.1038/nphoton.2015.201.

[5] R.A. Beth, Mechanical Detection and Measurement of the Angular Momentum of Light, Phys. Rev. 50 (1936) 115–125. https://doi.org/10.1103/PhysRev.50.115.

[6] L. Allen, M.W. Beijersbergen, R.J.C. Spreeuw, J.P. Woerdman, Orbital angular momentum of light and the transformation of Laguerre-Gaussian laser modes, Phys. Rev. A 45 (1992) 8185–8189. https://doi.org/10.1103/PhysRevA.45.8185.

[7] G. Gibson, J. Courtial, M.J. Padgett, M. Vasnetsov, V. Pas'ko, S.M. Barnett, S. Franke-Arnold, Free-space information transfer using light beams carrying orbital angular momentum, Opt. Express 12 (2004) 5448. https://doi.org/10.1364/OPEX.12.005448.

[8] G. Vallone, V. D'Ambrosio, A. Sponselli, S. Slussarenko, L. Marrucci, F. Sciarrino, P. Villoresi, Free-Space Quantum Key Distribution by Rotation-Invariant Twisted Photons, Phys. Rev. Lett. 113 (2014) 060503. https://doi.org/10.1103/PhysRevLett.113.060503.

[9] F. Bouchard, R. Fickler, R.W. Boyd, E. Karimi, High-dimensional quantum cloning and applications to quantum hacking, Sci. Adv. 3 (2017) e1601915. https://doi.org/10.1126/sciadv.1601915.

[10] J. Wang, J.-Y. Yang, I.M. Fazal, N. Ahmed, Y. Yan, H. Huang, Y. Ren, Y. Yue, S. Dolinar, M. Tur, A.E. Willner, Terabit free-space data transmission employing orbital angular momentum multiplexing, Nature Photon 6 (2012) 488–496. https://doi.org/10.1038/nphoton.2012.138.

[11] A.T. O'Neil, I. MacVicar, L. Allen, M.J. Padgett, Intrinsic and Extrinsic Nature of the Orbital Angular Momentum of a Light Beam, Phys. Rev. Lett. 88 (2002) 053601. https://doi.org/10.1103/PhysRevLett.88.053601.

[12] K.I. Willig, S.O. Rizzoli, V. Westphal, R. Jahn, S.W. Hell, STED microscopy reveals that synaptotagmin remains clustered after synaptic vesicle exocytosis, Nature 440 (2006) 935–939. https://doi.org/10.1038/nature04592.

[13] E. Hemsing, A. Knyazik, M. Dunning, D. Xiang, A. Marinelli, C. Hast, J.B. Rosenzweig, Coherent optical vortices from relativistic electron beams, Nature Phys 9 (2013) 549–553. https://doi.org/10.1038/nphys2712.

[14] E. Hemsing, A. Marinelli, J.B. Rosenzweig, Generating Optical Orbital Angular Momentum in a High-Gain Free-Electron Laser at the First Harmonic, Phys. Rev. Lett. 106 (2011) 164803. https://doi.org/10.1103/PhysRevLett.106.164803.

[15] U.D. Jentschura, V.G. Serbo, Generation of High-Energy Photons with Large Orbital Angular Momentum by Compton Backscattering, Phys. Rev. Lett. 106 (2011) 013001. https://doi.org/10.1103/PhysRevLett.106.013001.

[16] C. Hernández-García, A. Picón, J. San Román, L. Plaja, Attosecond Extreme Ultraviolet Vortices from High-Order Harmonic Generation, Phys. Rev. Lett. 111 (2013) 083602. https://doi.org/10.1103/PhysRevLett.111.083602.

[17] K.M. Dorney, L. Rego, N.J. Brooks, J. San Román, C.-T. Liao, J.L. Ellis, D. Zusin, C. Gentry, Q.L. Nguyen, J.M. Shaw, A. Picón, L. Plaja, H.C. Kapteyn, M.M. Murnane, C. Hernández-García, Controlling the polarization and vortex charge of attosecond high-harmonic beams via simultaneous spin–orbit momentum conservation, Nature Photon 13 (2019) 123–130. https://doi.org/10.1038/s41566-018-0304-3.

[18] M. Zürch, C. Kern, P. Hansinger, A. Dreischuh, Ch. Spielmann, Strong-field physics with singular light beams, Nature Phys 8 (2012) 743–746. https://doi.org/10.1038/nphys2397.

[19] M. Padgett, R. Bowman, Tweezers with a twist, Nature Photon 5 (2011) 343–348. https://doi.org/10.1038/nphoton.2011.81.

[20] A. Leblanc, A. Denoeud, L. Chopineau, G. Mennerat, Ph. Martin, F. Quéré, Plasma holograms for ultrahigh-intensity optics, Nature Phys 13 (2017) 440–443. https://doi.org/10.1038/nphys4007.

[21] A. Denoeud, L. Chopineau, A. Leblanc, F. Quéré, Interaction of Ultraintense Laser Vortices with Plasma Mirrors, Phys. Rev. Lett. 118 (2017) 033902. https://doi.org/10.1103/PhysRevLett.118.033902.

[22] J. Vieira, R.M.G.M. Trines, E.P. Alves, R.A. Fonseca, J.T. Mendonça, R. Bingham, P. Norreys, L.O. Silva, Amplification and generation of ultra-intense twisted laser pulses via stimulated Raman scattering, Nat Commun 7 (2016) 10371. https://doi.org/10.1038/ncomms10371.

[23] J.W. Wang, M. Zepf, S.G. Rykovanov, Intense attosecond pulses carrying orbital angular momentum using laser plasma interactions, Nat Commun 10 (2019) 5554. https://doi.org/10.1038/s41467-019-13357-1.

[24] L. Zhang, B. Shen, Z. Bu, X. Zhang, L. Ji, S. Huang, M. Xiriai, Z. Xu, C. Liu, Z. Xu, Vortex Harmonic Generation by Circularly Polarized Gaussian Beam Interacting with Tilted Target, Phys. Rev. Applied 16 (2021) 014065. https://doi.org/10.1103/PhysRevApplied.16.014065.

[25] S.V. Bulanov, N.M. Naumova, F. Pegoraro, Interaction of an ultrashort, relativistically strong laser pulse with an overdense plasma, Physics of Plasmas 1 (1994) 745–757. https://doi.org/10.1063/1.870766.

[26] R. Lichters, J. Meyer-ter-Vehn, A. Pukhov, Short-pulse laser harmonics from oscillating plasma surfaces driven at relativistic intensity, Physics of Plasmas 3 (1996) 3425–3437. https://doi.org/10.1063/1.871619.

[27] T. Baeva, S. Gordienko, A. Pukhov, Theory of high-order harmonic generation in relativistic laser interaction with overdense plasma, Phys. Rev. E 74 (2006) 046404. https://doi.org/10.1103/PhysRevE.74.046404.

[28] Z.-Y. Chen, A. Pukhov, Bright high-order harmonic generation with controllable polarization from a relativistic plasma mirror, Nat Commun 7 (2016) 12515. https://doi.org/10.1038/ncomms12515.

[29] W.P. Wang, C. Jiang, B.F. Shen, F. Yuan, Z.M. Gan, H. Zhang, S.H. Zhai, Z.Z. Xu, New Optical Manipulation of Relativistic Vortex Cutter, Phys. Rev. Lett. 122 (2019) 024801. https://doi.org/10.1103/PhysRevLett.122.024801.

[30] K. Toyoda, F. Takahashi, S. Takizawa, Y. Tokizane, K. Miyamoto, R. Morita, T. Omatsu, Transfer of Light Helicity to Nanostructures, Phys. Rev. Lett. 110 (2013) 143603.

https://doi.org/10.1103/PhysRevLett.110.143603.

[31] K. Toyoda, K. Miyamoto, N. Aoki, R. Morita, T. Omatsu, Using Optical Vortex To Control the Chirality of Twisted Metal Nanostructures, Nano Lett. 12 (2012) 3645–3649. https://doi.org/10.1021/nl301347j.

[32] L. Yi, Anomalous Effects in Single-slit Diffraction of Light at Relativistic Intensities, (2025). https://doi.org/10.48550/arXiv.2505.03199.

[33] K. Hu, X. Guo, L. Yi, High-Harmonic Generation and Optical Torque Interaction via Relativistic Diffraction of a Spatiotemporal Vortex Light, (2025). https://doi.org/10.48550/arXiv.2505.03215.

[34] Y. Meng, R. Li, L. Yi, Relativistic Oscillating Window Driven by an Intense Laguerre Gaussian Laser Pulse, (2025). https://doi.org/10.48550/arXiv.2506.21407.

[35] T.D. Arber, K. Bennett, C.S. Brady, A. Lawrence-Douglas, M.G. Ramsay, N.J. Sircombe, P. Gillies, R.G. Evans, H. Schmitz, A.R. Bell, C.P. Ridgers, Contemporary particle-in-cell approach to laser-plasma modelling, Plasma Phys. Control. Fusion 57 (2015) 113001. https://doi.org/10.1088/0741-3335/57/11/113001.

[36] V.G. Niziev, A. Nesterov-Mueller, Divergence features of laser beams with angular momentum, Quantum Electron. 51 (2021) 1122–1126. https://doi.org/10.1070/QEL17658.

[37] K. Hu, L. Yi, Intense high-harmonic optical vortices generated from a microplasma waveguide irradiated by a circularly polarized laser pulse, Phys. Rev. Research 4 (2022) 033095. https://doi.org/10.1103/PhysRevResearch.4.033095.

[38] G. Doumy, F. Quéré, O. Gobert, M. Perdrix, Ph. Martin, P. Audebert, J.C. Gauthier, J.-P. Geindre, T. Wittmann, Complete characterization of a plasma mirror for the production of high-contrast ultraintense laser pulses, Phys. Rev. E 69 (2004) 026402. https://doi.org/10.1103/PhysRevE.69.026402.

[39] U. Teubner, U. Wagner, E. Förster, Sub-10 fs gating of optical pulses, J. Phys. B: At. Mol. Opt. Phys. 34 (2001) 2993–3002. https://doi.org/10.1088/0953-4075/34/15/306.

[40] B. Dromey, S. Kar, M. Zepf, P. Foster, The plasma mirror—A subpicosecond optical switch for ultrahigh power lasers, Review of Scientific Instruments 75 (2004) 645–649. https://doi.org/10.1063/1.1646737.

[41] N.K. Fontaine, R. Ryf, H. Chen, D.T. Neilson, K. Kim, J. Carpenter, Laguerre-Gaussian mode sorter, Nat Commun 10 (2019) 1865. https://doi.org/10.1038/s41467-019-09840-4.

[42] S. Lightman, G. Hurvitz, R. Gvishi, A. Arie, Miniature wide-spectrum mode sorter for vortex beams produced by 3D laser printing, Optica 4 (2017) 605. https://doi.org/10.1364/OPTICA.4.000605.

[43] A.M. Yao, M.J. Padgett, Orbital angular momentum: origins, behavior and applications, Adv. Opt. Photon. 3 (2011) 161. https://doi.org/10.1364/AOP.3.000161.

[44] A. D'Errico, R. D'Amelio, B. Piccirillo, F. Cardano, L. Marrucci, Measuring the complex orbital angular momentum spectrum and spatial mode decomposition of structured light beams, Optica 4 (2017) 1350. https://doi.org/10.1364/OPTICA.4.001350.